\documentclass[12pt,a4paper]{article}
\usepackage[latin9]{inputenc}

\makeatletter

\pdfpageheight\paperheight
\pdfpagewidth\paperwidth

\usepackage{amsfonts}
\usepackage{hyperref}

\makeatother

\begin{document}

\title{Developments on the Jordan-Schwinger construction and contraction for the $su_q(2)$}

\author{R.\ Kullock%
\thanks{{\em e-mail: ricardokl@cbpf.br}%
} \\
\\
 \textit{$~^{\dagger}$Centro Brasileiro de Pesquisas F\'isicas} \\
\textit{Rio de Janeiro} \vspace{0.5cm}
}
\maketitle
\begin{abstract}
The contraction and Jordan-Schwinger construction connect the $su(2)$
and the heisenberg algebra, going in oposite directions. This persists
in the q-deformed cases, but in a slightly different way. This work
investigates this further, discussing some details and results found in
the litterature and presenting some new ones, including a nonlinear choice
for the contraction.

\vspace{0.2cm}

{\footnotesize { }\textbf{\footnotesize Keywords}{\footnotesize :
q-deformation, contraction, boson}{\footnotesize \par}

 }{\footnotesize \par}
\end{abstract}
\newpage{}

\section{Introduction}

In this work the relation between the $su(2)$ and the Heisenberg
$h(1)$ algebra will be investigated, with a focus in the q-deformed
cases $su_{q}(2)$ \cite{skly} and $h_{q}(1)$\cite{mac,kulish1}, using the Jordan-Schwinger representation
and a contraction procedure. For the deformed case, we will stress the diferent choices of deformation,
pointing how the same $su_{q}(2)$ is connected to different q-deformations of the Heisenberg algebra $h_{q}(1)$.
A nonlinear choice for the contraction is possible, taking the $su_{q}(2)$ the appropriate q-boson algebra.
Also, depending on the choice for the $su_{q}(2)$, the Hopf-algebra consistent conjugation does not survive
the contraction. In both cases, the representation gains a central focus.

Deformation theory in physics is often related to the idea of a minimal length \cite{heis,fr1,fr2},
in noncommutative quantum mechanics, or in quantum field theory \cite{szabo,gr1,gr2} with its recent interest being
string theory inspired \cite{sw}.
The noncommutativity may be introduced as Moyal products \cite{moy}, as a commutator ansatz \cite{snyder}, or perhaps
as a Drinfel'd twist \cite{cct,ccrt,crt}. More of less, it is usually desirable to have some Hopf algebra
behind it \cite{klym}.

Another application to deformations is the Yang-Baxter equation \cite{jimbo}, where q-deformations are used as a way
to derive its solutions. Another possibility is using the Drinfe'd twist \cite{drin}, which generates solutions from
a given one, including the trivial solution.

The q-deformation is among the most studied deformations, with the theory of the $su_q(2)$ serving as a model for other cases \cite{klym}.
The number $q$ is often taken to be real, but it can also be a phase (root of unity). When it is a real number, the representation
theory is completely analogous to the undeformed case, replacing coeficients by their q-deformed couterparts. In fact, the Hilbert
space representation has the same quantum numbers, and is isomorphic as a vector space to the undeformed case. Of course its realization
as a space of functions will be different.

Alongside with the $su_q(2)$, is the study of one-parameter deformations of the Heisenberg algebras. Of the many different
deformations, a few are called q-deformations, or q-bosons. They have their own applications, but they are also introduced
in connection to the $su_q(2)$, with the  JS (Jordan-Schwinger) construction \cite{mac} or the contraction proceedure \cite{kulish2}. 
These can be seen to be,
in some sense, inverses of each other in the undeformed case, but this does not seem to be true after the deformation. Instead of the 
JS contruction, it is also possible to use non-linear expressions.

These q-deformations can be thought as non-linear expressions of the original algebras. For the harmonic oscillator built using q-bosons, 
it is equivalent to an oscillator whose frequency depends on the initial data \cite{manko}. It is also possible to write the deformed rotation
algebra as a non-linear expression of the undeformed one \cite{zachos}.

When discussing q-deformed numbers, there are in fact two choices of deformation \cite{klym}. One is typacally used in the deformation of the $su(2)$,
while both show in the representation of q-boson algebras. Since the JS construction and contraction connect these algebras, they will
connect different q-numbers. In fact, as will be shown bellow, linear choices for the contraction cannot lead to the same q-number
deformation.

The contraction is more than just generating commutation relations, they make sense at the level of its representation \cite{madore,dooley,catt}. 
Often the parameter used for it is taken arbitrary, but it should be taken to be related to the dimension of the representation. Then, the 
contraction is defined at each step in a specific representation, leading eventually to an infinite tower of states. In this sense, the 
representation point of view of the contraction for the q-deformed case should be rather central.

This paper is structured as follows. First, we review the undeformed case, pointing out some important aspects of the representation. Second,
the canonical choices of q-deformation are introduced, and then the deformed case is discussed, including a nonlinear choice for the contraction.
After this we investigate a different attempt at
relating the $su_q(2)$ and a q-boson algebra. A second choice for the q-deformation of $su(2)$ is discussed. Finally, we make some comments on the
contraction of the JS construction using tensor products, and present the conclusion of the paper.

\section{Jordan-Schwinger and contraction for $q=1$}

Let us start with the Jordan-Schwinger construction. The idea is to
take a pair of Heisenberg algebras $[a_{i},a_{j}^{\dagger}]=\delta_{ij}$
and build quadractic expressions from them, in such a way to form
a Lie algebra. The usual case is the construction of the rotation
algebra, but it can also be used for the non-compact form, or even to
bigger algebras. For the rotation algebra, one usually defines 

\begin{equation}
J_{+}=a_{1}^{\dagger}a_{2},\qquad J_{-}=a_{1}a_{2}^{\dagger},\qquad H=\frac{1}{2}\left(a_{1}^{\dagger}a_{1}-a_{2}^{\dagger}a_{2}\right).
\label{JSundeformed}
\end{equation}
It is easy to check this leads to $J_{+}^{\dagger}=J_{-}$, $H^{\dagger}=H$,
and 

\begin{equation}
[H,J_{\pm}]=\pm J_{\pm},\qquad[J_{+},J_{-}]=2H.
\end{equation}

For the contraction \cite{madore}, we need to use a representation of the $su(2)$.
Take a vector $v_{k}$ such that $Hv_{k}=kv_{k}$, and $Cv_{k}=s(s+1)v_{k}$
is the Casimir, with $-s\geq k\geq s$. The contraction to the heisenberg
algebra is performed with the limit $s\rightarrow\infty$ for $k\geq0$
of the rescalings

\begin{equation}
\tilde{J_{\pm}}=\frac{J_{\pm}}{\sqrt{2s}},\qquad\tilde{H}=H+s,
\end{equation}
leading to the commutation relations

\begin{equation}
[\tilde{J_{-}},\tilde{J_{+}}]=1-\frac{\tilde{H}}{s},\qquad[\tilde{H},\tilde{J_{\pm}}]=\pm\tilde{J_{\pm}}.
\end{equation}

The limit of this commutators take $\tilde{J_{+}}$ and $\tilde{J_{-}}$
into the creation and aniquilation operators, and $\tilde{H}\rightarrow N$,
where $N$ is the number operator. It is important to notice the number
operator here is not equal to its usual quadratic expression, although
it is possible, by using the Casimir $C=\mathbf{J}^{2}$, to prove
it has the same action in the state vectors. These are defined by

\begin{equation}
|k\rangle=\lim_{s\rightarrow\infty}v_{k-s}
\end{equation} 
(and here it is obvious why the restriction $k\geq0$ is necessary). 

This construction works as well by taking the representation directly,
where

\begin{equation}
\tilde{J_{+}}v_{k-s}=\sqrt{-\frac{k(k+1)}{2s}+(k+1)}v_{(k+1)-s},\qquad\tilde{H}v_{k-s}=kv_{k-s},
\end{equation}
and similarly to $\tilde{J_{-}}$. Therefore, it is easy to see that
the limit takes the action of $\tilde{J}_{+}$ to its appropriate
one.

It should be stressed the imprtance of the shift in the vectors of
the representation, and the little consequence of the shift in $H$.
In fact, without using the redefined $H$ and using the appropriate
representation described above, we find

\begin{equation}
[\tilde{J_{-}},\tilde{J_{+}}]v_{k-s}=-\frac{H}{s}v_{k-s}=-\frac{k-s}{s}v_{k-s}
\end{equation}
and we have the correct limit. In the same way, had we chosen to shift
$H$ but not the vectors, we would find the commutator to be 

\begin{equation}
[\tilde{J_{-}},\tilde{J_{+}}]v_{k}=\left(1-\frac{\tilde{H}}{s}\right)v_{k}=\left(1-\frac{k+s}{s}\right)v_{k},
\end{equation}
and now the limit takes this into the translation algebra. This will
be further discussed bellow, using a tensor product representation
to the Jordan-Schwinger construction.

\section{The case for $q>1$}

\subsection{Definitions}
There are two different q-deformations of the $su(2)$ algebra,
and we will be dealing initially with just one of them,  as it is usually 
used for both the Jordan-Schwinger construction and the contraction. 
 Take the $su_{q}(2)$ given by the following

\begin{equation}
[H,J_{\pm}]=\pm J_{\pm},\qquad[J_{+},J_{-}]=[2H],
\end{equation}
where $[x]=\frac{q^{x}-q^{-x}}{q-q^{-1}}$ is the q-number (and this
can be extended to operators using formal power series).

For the Heisenberg algebra, there are also a few possibilities, but there
are two that are most relevant here. The first one, let us call it
$h_{q}^{1}$, is 

\begin{equation}
aa^{\dagger}-qa^{\dagger}a=q^{-N},\qquad aa^{\dagger}-q^{-1}a^{\dagger}a=q^{N},
\end{equation}
with $[N,a^{\dagger}]=a^{\dagger}$ and $[N,a]=-a$, and this leads
to the representation

\begin{equation}
a^{\dagger}|n\rangle=\sqrt{[n+1]}|n+1\rangle,\qquad a|n\rangle=\sqrt{[n]}|n-1\rangle,\qquad N|n\rangle=n|n\rangle,
\end{equation}
Notice in this case $a^{\dagger}a=[N]$ and $aa^{\dagger}=[N+1]$,
while this is not true had we chosen only the first relation for $a$
and $a^{\dagger}$. This other choice can be shown to be equivalent
to the present one only when its Casimir is zero, and will not lead
to a $su_{q}(2)$ otherwise. 

Another deformation, let us call this one $h_{q}^{2}$, is given by 

\begin{equation}
[A,A^{\dagger}]=q^{N},
\end{equation}
with $[N,A^{\dagger}]=A^{\dagger}$ and $[N,A]=-A$, and the representation 

\begin{equation}
A^{\dagger}|n\rangle=\sqrt{\frac{q^{n+1}-1}{q-1}}|n+1\rangle,\quad A|n\rangle=\sqrt{\frac{q^{n}-1}{q-1}}|n-1\rangle,\quad N|n\rangle=n|n\rangle,
\end{equation}
leading to another definition of q-numbers also found in the literature.

\subsection{Jordan-Schwinger construction and contraction}

The Jordan-Schwinger construction for the $su_{q}(2)$ introduced
above uses two copies of $h_{q}^{1}$, with similar definitions as
before,

\begin{equation}
J_{+}=a_{1}^{\dagger}a_{2},\qquad J_{-}=a_{1}a_{2}^{\dagger},\qquad H=\frac{1}{2}(N_{1}-N_{2}),
\end{equation}
keeping in mind here $N$ is a formal element of the algebra, and
not its usual quadractic expression. The construction may be explicitly verified
by using $[N_{1}][N_{2}+1]-[N_{1}+1][N_{2}]=[N_{1}-N_{2}]$, as seen in \cite{klym}.

For the contraction, we find a very clear distinction from the undeformed
case. Here, the natural generalization for the rescalings are given
by

\begin{equation}
\tilde{J_{\pm}}=\frac{J_{\pm}}{\sqrt{[2s]}},\qquad\tilde{H}=H+s,
\end{equation}
but this does not lead to $h_{q}^{1}$. Taking

\begin{equation}
\lim_{s\rightarrow\infty}\tilde{J}_{-}=b,\qquad\lim_{s\rightarrow\infty}\tilde{J}_{+}=b^{\dagger},\qquad\lim_{s\rightarrow\infty}\tilde{H}=N,
\end{equation}
for $q>1$ we will have

\begin{equation}
[b,b^{\dagger}]=q^{-2N},
\end{equation}
and the same shift in the eigenvectors can be used as in the undeformed
case. This result can be found in \cite{kulish2}. 
The reason for using the same shift will be clear in the discussion bellow,
but for now we can just say this shift in nothing more than choosing
an appropriate subspace. Notice as well that the discussion of the
importance of redefining $H$ remains the same here.

This is not the q-deformation as it is usually introduced, shown above,
and one usually uses a replacement $q\rightarrow q^{-1/2}$. Of course
this means the contraction limit is not the same algebra as before,
although they can be related by an appropriate redefinition of $b$
and $b^{\dagger}$.

This can be further understood by working directly on the representation.
Here we can see that

\begin{equation}
\tilde{J_{+}}v_{k-s}=\sqrt{[k+1]}\frac{\sqrt{[2s-k]}}{\sqrt{[2s]}}v_{(k+1)-s}.
\end{equation}
But the limit of $\sqrt{[2s-k]}/\sqrt{[2s]}$ is not trivial as before,
and we get $q^{-k/2}$ instead,
\begin{equation}
\lim_{s\rightarrow \infty} \tilde{J_{+}} v_{k-s}= \sqrt{[k+1]} q^{-k/2} |k-1\rangle.
\end{equation}

This is in fact the transformation needed on
the algebra mentioned before, where $b$ is reascaled as $bq^{N/2}$,
and here we see explicitly why it is needed. 

Instead of redefining the resulting ladder operators, we could change
the definition of $\tilde{J}_{\pm}$. The rescaling used before was
a natural generalization of the undeformed case, but there is no reason
why the definition shouldn't be nonlinear (in fact the algebras we
are dealing with are nonlinear themselves). By taking the outcome
of the contraction as seen in the representation, we have that the
choice

\begin{equation}
\tilde{J}_{+}=\frac{J_{+}}{\sqrt{[2s]}}q^{(H+s)/2}
\end{equation}
will lead directly into the correct action of $b^{\dagger}$. In fact,
this is just performing the transformation before the limit is taken.
This expression is no longer linear in the generators, but it leads
to the undeformed case when $q\rightarrow1$.

For this choice we may also find the commutators ($\tilde{J}_-=\tilde{J}_+^{\dagger}$)

\begin{equation}
 [\tilde{J}_-,\tilde{J}_+]=(q-1) \tilde{J}_+\tilde{J}_- + q^{\tilde{H}}\frac{[2H]}{[2s]}, 
\end{equation}
in other words,
\begin{equation}
 \tilde{J}_-\tilde{J}_+ - q \tilde{J}_+\tilde{J}_- = q^{\tilde{H}}\frac{[2H]}{[2s]},
\end{equation}
with limit
\begin{equation}
 b b^{\dagger}-qb^{\dagger} b = q^{-N}.
\end{equation}
Notice however we only find the first relation of the algebra for this choice of contraction. Nevertheless, since the contraction is built to make
sense in the representation, the second relation ($q\rightarrow q^{-1}$) will be valid as well.

Although $[n]$ is invariant for $q\rightarrow q^{-1}$, making this change in the definition of $\tilde{J}_{\pm}$ will not lead to 
$b b^{\dagger}-q^{-1}b^{\dagger} b = q^{N}$. This means the invariance exists before and after the contraction, although
the proceedure itself is not invariant.

\subsection{A different approach}

One issue with trying to have the same bosonic algebra in the Jordan-Schwinger
construction and the contraction limit is that we have used q-commutators
$[A,B]_{q}=AB-qBA$ on the initial algebra. It will be instructive
to use the formally equivalent

\begin{equation}
[a,a^{\dagger}]=[N+1]-[N]=\frac{q^{N}(q+1)-q^{-N}(q^{-1}+1)}{q-q^{-1}}.
\end{equation}
Now, we could look for some appropriate function $g(q,s)$ such that
$\tilde{J}_{\pm}=J_{\pm}/\sqrt{g}$ will lead to

\begin{equation}
\lim_{s\rightarrow\infty}[\tilde{J}_{-},\tilde{J}_{+}]=\frac{q^{N}(q+1)-q^{-N}(q^{-1}+1)}{q-q^{-1}},
\end{equation}
Looking for such $g(q,s)$ we arrive at the conclusion that its limit
is not singular. In fact we could just choose 

\begin{equation}
a^{\dagger}=\frac{(q+1)^{1/2}}{q^{1/4}}J_{-},\qquad a=\frac{(q+1)^{1/2}}{q^{1/4}}J_{+},\qquad2H=N+\frac{1}{2}.
\end{equation}
This leads to an isomorphism bettween the two algebras, inverting
the roles of the ladder operators. This is of course a surprise, considering
the two algebras lead to very different outcomes when $q\rightarrow1$,
and that the transformation is not singular in the same limit. We
could preserve the roles of the ladder operators, but only if $2H=-N-\frac{1}{2}$.

Considering our previous discussion, this isomorphism fails to be
true in each representation. Notice as well that we don't have the
possibility of choosing a shift in the basis, since we are no longer
considering a limit in $s$. Instead, what could be argued here is
that each of these algebras seems to admit a new representation, like
taking $H|k\rangle=(\frac{k}{2}+\frac{1}{4})|k\rangle$, and a finite
dimentional representation for the boson operators.

Since there is no limiting proceedure, this is not a contraction. There is a similar result
in the litterature, but again it should not be called a contraction.

\section{Another $su_q(2)$}

Just like there are many possible choices for the q-deformation of the Heisenberg algebra, there is also another choice for
the deformation of th $su(2)$. Although at the algebra level the change seems rather simple, the Hopf algebra it is somewhat
different, leading to some difficulties.

The commutators are given by

\begin{equation}
[H,J_{\pm}]=\pm 2 J_{\pm},\qquad[J_{+},J_{-}]=[H].
\end{equation}

One might be tempted to define $ \tilde{J_{\pm}} = J_{\pm}/\sqrt{[s]} $ leading to the limit $ [b,b^{\dagger}]=q^{-N} $ but this not so.
It should be kept in mind the discussion of how the representation plays an important role in the contraction, and here 
$H v_k = 2k v_k$. Therefore, the definitions for $\tilde{J}_{\pm}$ will be the same as before,

\begin{equation}
 \tilde{J}_{\pm} = \frac{J_{\pm}}{\sqrt{[2s]}},
\end{equation}
but now we have
\begin{equation}
 \tilde{H} = H + 2s,
\end{equation}
and the limit leads reather to $N |n \rangle = 2n |n \rangle$. The state vectors are defined as before, with the same shift.
Had we chosen the naive rescaling, the commutator of the ladder operators would be ill-defined.

The less than trivial issue will be the conjugation. As a consequence of its Hopf algebra, the star structure here is given
by $H^{\dagger}=H$ and
\begin{equation}
 J_+^{\dagger} = J_- q^{H}, \qquad J_-^{\dagger} = q^{-H} J_+,
\end{equation}
so that $(J_{\pm}^{\dagger})^{\dagger}=J_{\pm}$, making it involutive and with the correct limit in $q\rightarrow 1$. Dividing
both sides by $\sqrt{[2s]}$, we have
\begin{equation}
 \tilde{J}_+^{\dagger} = \tilde{J}_- q^{\tilde{H}-2s}, \qquad \tilde{J}_-^{\dagger} = q^{-\tilde{H}+2s} \tilde{J}_+,
\end{equation}
so that the first one goes to zero, while the second diverges (for $q>1$), and the conjugation structure will not be carried 
over to the contraction limit, although the action of the ladder operators will be the same as before.

\section{Tensor products}

Further insight may be gained by considering the contraction limit of the tensor
product description of the Jordan-Schwinger representation. We have that $v_{k}=|s+k\rangle\otimes|s-k\rangle$, and therefore $v_{k-s}=|k\rangle\otimes|2s-k\rangle$.
With this, 

\begin{equation}
b^{\dagger}=\lim_{s\rightarrow\infty}a^{\dagger}|k\rangle\otimes\frac{a}{\sqrt{2s}}|2s-k\rangle,
\end{equation}
and it becomes obvious the limit simply disapears with the right hand
side of the tensor product. The coeficient of $a/\sqrt{2s}$ goes
to unity and the state vector is taken to infinity. With $b$ it is
the same, although with the number operator both sides are necessary,
with

\begin{equation}
N|k\rangle=\lim_{s\rightarrow\infty}
\left(\frac{a^{\dagger}a}{2}|k\rangle\otimes|2s-k\rangle-|k\rangle\otimes \frac{a^{\dagger}a}{2}|2s-k\rangle+
s|k\rangle\otimes|2s-k\rangle\right),
\end{equation}
so that $N|k\rangle=k\lim_{s\rightarrow\infty}\left(|k\rangle\otimes|2s-k\rangle\right)$,
taking the right hand side state vector to infinity.

Initially, the $su(2)$ acts on a vector space spanned by state vectors
$|0\rangle\otimes|2s\rangle$, when $k=-s$, up to $|2s\rangle\otimes|0\rangle$,
when $k=s$, so that on the Jordan-Schwinger construction only
some of the state vectors of the boson representation are used. The
shift on $v_{k}$ allows only for $k\geq0$ (otherwise the representation
would be ill-defined), and it picks a subspace of the representation
$v_{-s},\cdots v_{0}$. As $s\rightarrow\infty$, this becomes a full
tower of states, but upside down, with the vacuum being at infinity,
which of course makes much more sense in the tensor product described
above.
For the $q$-deformed case, we have
\begin{equation}
J_{+}|k\rangle\otimes|2s-k\rangle=\sqrt{[k+1]}|k+1\rangle\otimes\sqrt{[2s-k]}|2s-k-1\rangle
\end{equation}
so that it is precisely the right hand side of the tensor product that fails to go to unity upon rescaling and contraction. The solution for this is described above.

\section{Conclusion}

In this work we discuss the JS construction and the contraction of q-deformed $su(2)$ and Heisenberg algebras. Some focus is given
to the representation aspects, and also to an isomorphism between the two algebras, which in fact fails at the representation level.

When looking for the appropriate definitions for the contraction, the choice for $H$ is of little consequence, as far as the commutation relations go.
It will indeed allow for the correct limit into the number operator, but the shift in the representation vector will play a more central role.

Dealing with the deformed version of the JS construction and contraction, we are faced with a few different choices, even with different definitions
of q-numbers at the representation level. This means the two are no longer the inverse of one another. As was mentioned before, this can be solved
by using nonlinear definitions for the contraction. This leads to the correct representation but only one of the q-boson relations is achieved. The symmetry
$q \rightarrow q^{-1}$ exists both before and after the contraction, but the nonlinear expression used does not have it. As far as the second choice for the deformation of the $su(2)$,
it is still possible to define the contraction, but its conjugation structure will no survive.

Trying to find a contraction leading back into the original q-boson in the q-deformed JS construction, we are lead to what seems to be an isomorphism.
It should be noted that this is not in fact a contraction, since no limit is taken. In fact, it will not hold at the representation level.

Looking into the contraction of a JS concstruction using tensor products can be insighful. We see the contraction of the ladder operators in the
undeformed case is just keeping the left hand side of the tensor product, taking the right hand side coeficient to one, and the state vector 
to infinity. With this approach we can see there is little room for different choices (even inverting the roles of the left and right hand sides
is not possible).

{}~ \\
{}~ 

\textbf{\large Acknowledgments}{} ~\\
{}~

This work was supported by CNPq -- Brasil. The author is thankful for the comments from F. Toppan.

\end{document}